\documentclass[prx,letterpaper,nobalancelastpage,twocolumn,superscriptaddress,nofootinbib,longbibliography]{revtex4-2}

\usepackage{graphicx}
\usepackage{amsmath}
\usepackage{bbold}
\usepackage{amssymb}
\usepackage[english]{babel}
\usepackage{color}
\usepackage[version=4]{mhchem}
\usepackage[hidelinks]{hyperref}
\usepackage{dsfont}
\usepackage{placeins}
\usepackage{mathtools}
\usepackage[normalem]{ulem}
\usepackage{lipsum}

\setcitestyle{super}

\usepackage{soul}
\newcommand{\Caltech}{California Institute of Technology, Pasadena, CA 91125, USA}

\newcommand{\Stanford}{Department of Electrical Engineering, Stanford University, Stanford, CA, USA}

\usepackage{caption}

\DeclareCaptionLabelSeparator{bar}{ \textbf{\textbar}~}
\usepackage{enumitem}
\setlist{nolistsep}

\begin{document}

\title{Multi-ensemble metrology by programming local rotations with atom movements}
\author{Adam L. Shaw}\thanks{These authors contributed equally to this work}
\author{Ran Finkelstein}\thanks{These authors contributed equally to this work}
\author{Richard Bing-Shiun Tsai}
\author{Pascal Scholl}
\affiliation{\Caltech}
\author{\\Tai Hyun Yoon}\thanks{Permanent address: Department of Physics, Korea University, Seoul 02841, Republic of Korea}
\affiliation{\Caltech}
\author{Joonhee Choi}
\affiliation{\Caltech}
\affiliation{\Stanford}
\author{Manuel Endres}\email{mendres@caltech.edu}
\affiliation{\Caltech}

\maketitle
\textbf{Current optical atomic clocks do not utilize their resources optimally. In particular, an exponential gain in sensitivity could be achieved if multiple atomic ensembles were to be controlled or read-out individually, even without entanglement. However, controlling optical transitions locally remains an outstanding challenge for neutral atom based clocks and quantum computing platforms. Here we show arbitrary, single-site addressing for an optical transition via sub-wavelength controlled moves of tweezer-trapped atoms, which we perform with $99.84(5)\%$ fidelity and with $0.1(2)\%$ crosstalk to non-addressed atoms. The scheme is highly robust as it relies only on relative position changes of tweezers and requires no additional addressing beams. Using this technique, we implement single-shot, dual-quadrature readout of Ramsey interferometry using two atomic ensembles simultaneously, and show an enhancement of the usable interrogation time at a given phase-slip error probability. Finally, we program a sequence which performs local dynamical decoupling during Ramsey evolution to evolve three ensembles with variable phase sensitivities, a key ingredient of optimal clock interrogation. Our results demonstrate the potential of fully programmable quantum optical clocks even without entanglement and could be combined with metrologically useful entangled states in the future.}

Sensors based on quantum probes provide some of the most precise measurements in science~\cite{Ludlow2015,Safronova2018,Andreev2018,Roussy2022,Degen2017}. For many such systems, fundamental sensitivity limits can be improved through entanglement~\cite{Pezze2018,Pedrozo2020,Macieszczak2014,Kaubruegger2021}, but in the presence of noise, a practical advantage of such schemes is not guaranteed~\cite{Huelga1997,Schulte2020}. A complementary approach studies optimal metrology with entanglement-free quantum control and readout methods. For both approaches, an important figure of merit is not just the sensitivity to a given observable, but also the dynamic range over which that observable can unambiguously be estimated~\cite{Rosenband2013,Borregaard2013,Colombo2022a,Demkowicz2012}.

In the particular case of optical atomic clocks~\cite{Ludlow2015}, the observable of interest is the stochastically evolving phase of a laser acting as a local oscillator, which is mapped into population imbalance of an ultra-narrow optical transition. The clock stability improves with the interrogation time, but the phase can only be unambiguously mapped when it is in the range $[-\pi/2, \pi/2]$; phases outside of this range lead to phase-slip errors, which limits the attainable interrogation time at a given phase-slip error probability in the case of local oscillator limited clocks. Optimal readout schemes~\cite{Rosenband2013,Borregaard2013,Li2022} could improve the attainable interrogation time exponentially but require local rotational control over sub-ensembles during the sensing protocol or local mid-circuit readout and reset, both of which have not been demonstrated to date.

\begin{figure*}[ht!]
	\centering
	\includegraphics[width=\textwidth]{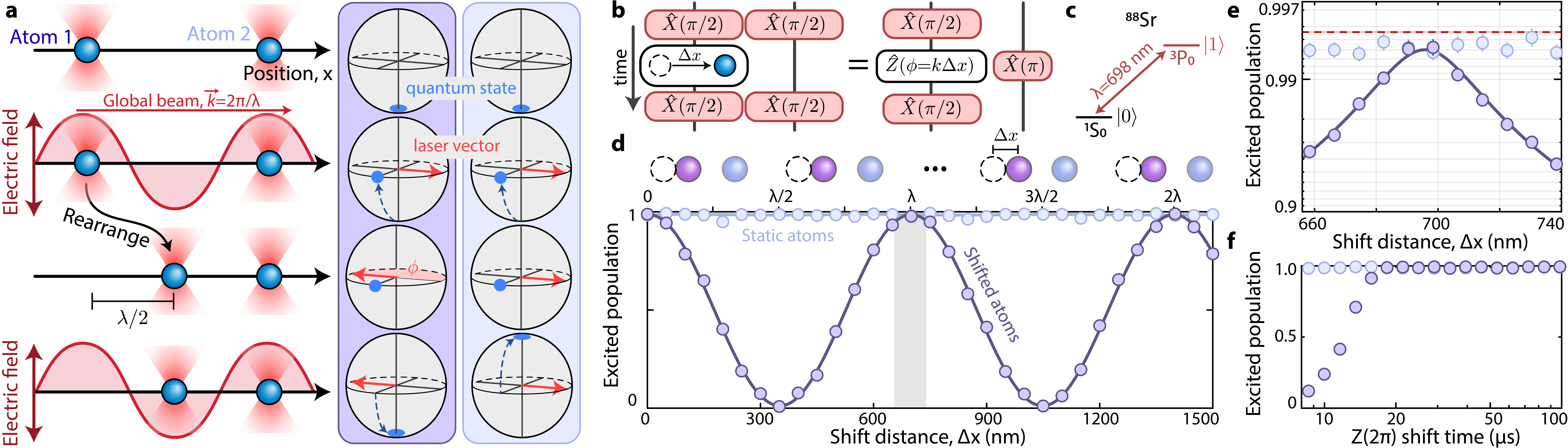}
	\caption{\textbf{Single-site addressing with movement-induced phase shifts.} \textbf{(a)} We consider two atoms individually trapped in optical tweezers, both initially in the electronic ground state. Traveling light emitted from a global laser beam applies a $\pi$/2 rotation to both atoms, and is disabled, but remains phase coherent with the atomic transition.  One of the atoms is then moved by half the laser wavelength, $\lambda$, from its initial position, rotating the effective local laser frame by an angle $\phi=\pi$. When the laser drive is restarted to apply another $\pi$/2 pulse, the moved atom now rotates back to the ground state, while the static atom rotates to the excited state. \textbf{(b)} Control over the atom displacement, $\Delta x$, is equivalent to arbitrary local rotations of the laser drive by $\phi=k\Delta x$ about the $\hat{Z}$-axis. \textbf{(c)} We implement this protocol with an array of $^{88}$Sr atoms utilizing the ultra-narrow $^1S_0{\leftrightarrow}^3P_0$ transition with $\lambda=698.4$ nm for global driving. \textbf{(d)} Top: with an array of 39 tweezers in one dimension, we apply the protocol in \textbf{b}, shifting every odd site (purple markers) in the array while leaving all even sites static (blue markers) during the dynamics. Bottom: a sinusoidal oscillation emerges in the excited state population of the shifted sites, with a period of 699(1) nm. \textbf{(e)} Focusing on the region around $\Delta x=\lambda$ (grey shaded region in \textbf{d}), we find the shifted atom shows no measurable loss in fidelity compared to the unshifted atoms. Correcting for the bare fidelity for performing a global $\hat{X}(\pi)$ rotation (red dashed line, 0.9956(1)), we find the shift operation is performed with a fidelity of 0.9984(5). The ratio of the shifted to unshifted fidelities is 0.9998(5), suggesting that the dominant source of error comes from global laser phase noise during the finite wait time required to perform the shift, rather than the movement itself. From data in \textbf{d}, we find the crosstalk to the static atoms is $0.1(2)\%$, consistent with 0. \textbf{(f)} The shift to apply a $\hat{Z}(2\pi)$ rotation can be performed without noticeable loss of fidelity down to shift times of ${\sim}20$ $\mu$s; data in $\textbf{e}$ are taken with a shift time of 32 $\mu$s, in addition to an extra wait time of 34 $\mu$s to account for finite jitter in the control timings.
 }  
	\vspace{-0.5cm}
	\label{Fig1}
\end{figure*}

Here we show local control of optical transitions in a tweezer array clock~\cite{Madjarov2019,Norcia2019a,Young2020} by using rearrangement techniques~\cite{Endres2016,Barredo2016,Bluvstein2022,Dordevic2021,Lengwenus2010,Beugnon2007} on atoms in superposition states to precisely control the position-dependent phase imprinted by light-matter interaction. The scheme~\cite{Schaetz2004,Chen2022} is experimentally simple and highly robust as it relies solely on the relative stability of tweezer positions and does not involve any auxiliary addressing beams. Using this technique, we demonstrate arbitrary, parallel, single-site-resolved optical qubit rotations with high fidelity.

We utilize such rotations to double the dynamic range of optical Ramsey spectroscopy by performing simultaneous evolution on two separate atomic ensembles within one tweezer array, each of which measures a different phase quadrature~\cite{Li2022}; we extend the coherent interrogation time by a factor of 3.43(13) relative to the standard, single ensemble sequence. Finally, we realize a proof-of-principle protocol for programming local dynamical decoupling sequences during Ramsey interrogation such that different ensembles within a single atom array have different sensitivities to phase variation, and discuss its implementation as part of a general protocol for improving clock stability~\cite{Rosenband2013,Borregaard2013}.

Aside from clocks, our technique for implementing local, parallel rotations about arbitrary axes might also find use in neutral atom quantum computing platforms utilizing optical transitions~\cite{Chen2022,Wu2022}, where local coherent control of optical qubits has not been demonstrated before. More generally, our results point to a future of fully programmable neutral atom optical clocks that incorporate features of quantum computers.

The basic principle of our scheme is illustrated in Fig.~\ref{Fig1}a. We consider two atoms both initially in the ground electronic state, $|0\rangle$, interacting with a global laser beam characterized by wavevector $k=2\pi/\lambda$, and wavelength $\lambda$, propagating along the array axis. With the globally applied laser, we create an equal superposition state of $|0\rangle$ and the excited state, $|1\rangle$; in a Bloch sphere picture, this corresponds to a $\pi/2$ rotation around the $x$-axis ($\hat{X}(\pi/2$)). The laser beam is then extinguished with an optical modulator, but remains phase coherent with the atomic transition. Using atom rearrangement techniques~\cite{Endres2016,Barredo2016,Bluvstein2022,Dordevic2021,Lengwenus2010,Beugnon2007}, one of the atoms is shifted from its original position by $\Delta x$, applying an effective phase shift of $\phi=k\Delta x$ (Methods). In Fig.~\ref{Fig1}a, we first consider the special case of $\Delta x=\lambda/2$, or equivalently a $\pi$ rotation around the $z$-axis ($\hat{Z}(\pi$)) for the shifted atom (Fig.~\ref{Fig1}a). Subsequently, we apply a second global $\hat{X}(\pi/2)$ rotation with the same laser as before; the shifted atom now rotates back to $|0\rangle$ because of the movement-induced phase shift, while the unmoved atom completes its rotation to $|1\rangle$. 

The main principle behind this scheme is a locally controlled change of the relative phase between the atomic dipole-oscillation and the phase of the laser while the atom is in a superposition state; in essence, our scheme realizes a locally controlled Ramsey sequence with global driving (Methods). Similar techniques have been used in the context of ion trap experiments with two ions~\cite{Schaetz2004}, but not in a scalable fashion, as is possible with tweezer arrays~\cite{Chen2022}.

\begin{figure*}[t!]
	\centering
	\includegraphics[width=\textwidth]{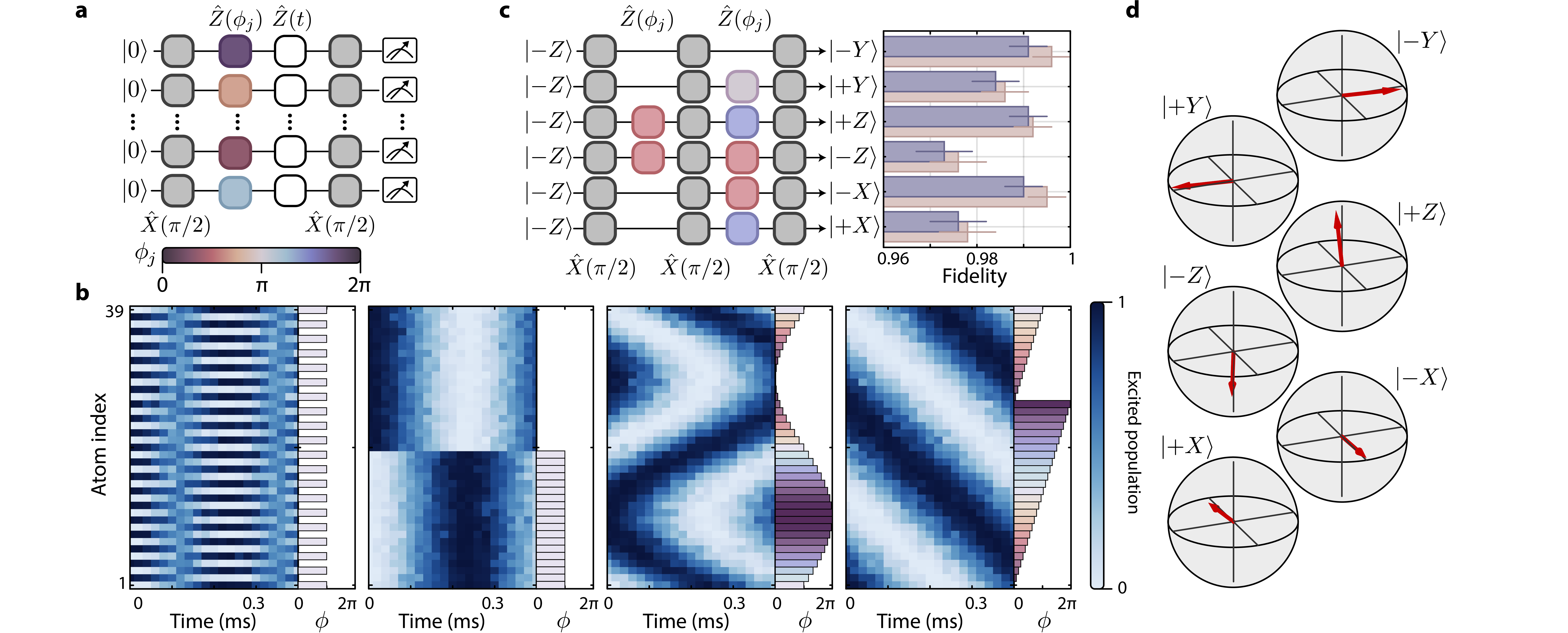}
	\caption{\textbf{Arbitrary, parallel, local rotations.} \textbf{(a)} We implement site-resolved phase shifts, $\phi_j$, during the dark time, $t$, of standard Ramsey interrogation by inserting arbitrary and parallel shifts of various distances to the array of atoms. \textbf{(b)} Results of this operation as a function of Ramsey time ($x$-axis) for different tweezers in the array ($y$-axis). The corresponding programmed phase-shift pattern is shown on the right of each subfigure. \textbf{(c)} By applying multiple global $\hat{X}(\pi/2)$ pulses (grey blocks), in tandem with local movement shifts (same color scale as in $\textbf{a}$), arbitrary local rotations can be performed. We show a demonstration by rotating an array of six atoms, initially in the $|0\rangle=|{-}Z\rangle$ state, in parallel to the six cardinal states ($|{-}Z\rangle, |{+}Z\rangle, |{-}Y\rangle, |{+}Y\rangle, |{-}X\rangle, |{+}X\rangle$), achieving an average fidelity of 0.984(2) (blue bars), and 0.987(2) after state preparation and measurement (SPAM) correction (tan bars), limited by global $\hat{X}(\pi/2)$ fidelity and decoherence during the time needed for movement (Methods). \textbf{(d)} Bloch sphere visualizations of the states measured with quantum state tomography in \textbf{c}.
	}
	\vspace{-0.5cm}
	\label{Fig2}
\end{figure*}

We show an experimental demonstration with our $^{88}$Sr optical tweezer array experiment~\cite{Cooper2018,Madjarov2019,Choi2023}. We employ a one-dimensional array of 39 optical tweezers generated via an acousto-optic deflector (AOD) driven by an arbitrary waveform generator (AWG). This allows for precise control over the relative tweezer positions at the nanometer level, enabling arbitrary $\hat{Z}(\phi)$ rotations (Fig.~\ref{Fig1}b). Global driving is performed on the ultra-narrow $^1S_0{\leftrightarrow}^3P_0$ optical clock transition with a transition wavelength of $\lambda=698.4$ nm (Fig.~\ref{Fig1}c).

In a first experiment, we apply an $\hat{X}(\pi/2)$ operation globally to the entire array, then shift every odd site by the same distance, $\Delta x$, apply another global $\hat{X}(\pi/2)$ rotation, and finally measure the excited state population in both the shifted and unshifted sub-arrays. The excited state population of shifted atoms, $P_s$, shows sinusoidal oscillations with a period of 699(1) nm as a function of $\Delta x$, consistent with $\phi/2\pi=\Delta x/\lambda$, where $\lambda$ is the transition wavelength. The quoted error on this measurement is purely statistical, and ignores potential systematic error arising from the independent distance calibration performed with an optical resolution test target. We note that the present measurement is likely a far more precise and accurate distance calibration tool, and could find use as an effective \textit{in-situ} laser-based ruler with applications for precision determination of distance-dependent inter-atom effects, such as Rydberg interactions~\cite{Beguin2013}.

To quantify the phase shift fidelity, we focus on a narrow region around $\Delta x= \lambda$, corresponding to a $\hat{Z}(2\pi)$ rotation (Fig.~\ref{Fig1}e). A quadratic fit to $P_s(\Delta x)$ shows a maximum value of $P_s=0.9940(5)$ (not corrected for state preparation and measurement (SPAM) errors), consistent with the mean excited state population of unshifted atoms, $P_u=0.9942(2)$, in the same range. Correcting for the bare $\hat{X}(\pi)$ fidelity (red dashed line) of 0.9956(1)  shows the shift operation is performed with a fidelity of 0.9984(5). We note that applying SPAM correction on the bare fidelities maintains the shift fidelity largely unchanged as it is calculated from the ratio of the two.  The ratio of the shifted to unshifted fidelities is 0.9998(5), suggesting that the dominant source of error comes from global laser phase noise during the finite wait time required to perform the shift, rather than the movement itself. We study the fidelity to perform the $\hat{Z}(2\pi)$ rotation as a function of the shift time (Fig.~\ref{Fig1}f), and find that the fidelity remains constant down to shift times of $t_s=20\ \mu$s; data in Fig.~\ref{Fig1}e were taken with $t_s=32\ \mu$s, plus an additional 34 $\mu$s of wait time to account for jitter in the subsequent control timings. Importantly, for all shift distances in Fig.~\ref{Fig1}d, the excited state population of the neighboring unshifted atoms is nearly constant, showing crosstalk of only $0.1(2)\%$ (Methods).

\begin{figure*}[ht!]
	\centering
	\includegraphics[width=\textwidth]{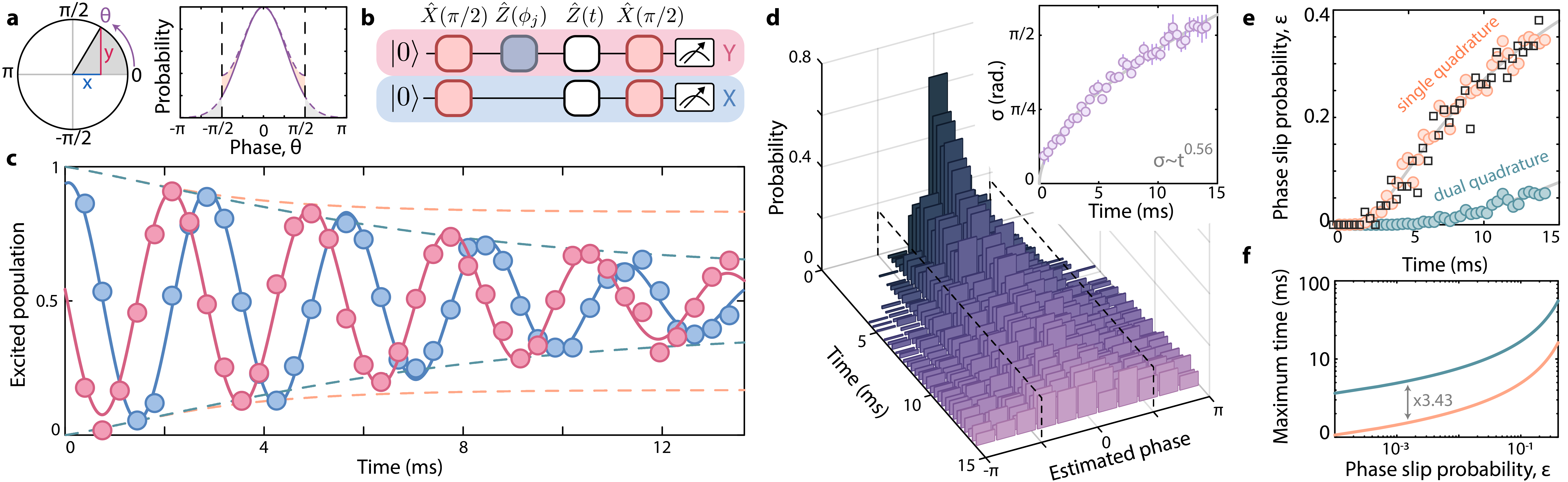}
	\caption{\textbf{Enhanced sensing with dual-quadrature measurement.} \textbf{(a)} For a given phase angle, $\theta$, population measurement in only a single basis, e.g. $Y$, can only be inverted within a dynamic range of $-\pi/2<\theta<\pi/2$.  By measuring both quadratures, $X$ and $Y$, this dynamic range can be doubled to $-\pi<\theta<\pi$, allowing for interrogating larger spreads in phase, such as when measuring for longer times. \textbf{(b)} We implement dual-quadrature readout of Ramsey interrogation by applying local $\pi/2$ phase shifts to all odd sites in the array. \textbf{(c)} With single-quadrature readout, the interrogation time is limited due to phase-slips, visible by the separation between a decay envelope reconstructed from the single-quadrature phase spread (orange dashed line) and the averaged Ramsey signal (blue and red markers and lines). The equivalent reconstruction with dual-quadrature readout (green dashed line) is accurate up to longer times. \textbf{(d)} To perform this reconstruction, we measure time-resolved probability distributions of the estimated phase relative to the mean from dual-quadrature measurement. As the standard deviation, $\sigma$, of the phase distribution grows (inset), the estimated phase begins exceeding the $-\pi/2<\theta<\pi/2$ range for normal spectroscopy (black dashed lines), but is still resolvable via dual-quadrature measurement. Note that the time-dependent contribution from quantum projection noise to the standard deviation has been subtracted off in the inset (Methods).  \textbf{(e)} We estimate the phase-slip probability, $\epsilon$, for single- (orange circles) and dual-quadrature (green circles) measurements by fitting a folded Gaussian to the time-resolved estimated phases in \textbf{d}. The fit is folded over at the boundaries of the dynamic range to account for the behavior of phase-slips, as in \textbf{a}. For the single-quadrature case, we also estimate the probability directly from the underlying data (squares), which is in good agreement with the estimate from fit. Solid lines are the predicted phase slip probabilities from the fit in the inset of \textbf{d}. This fit is used to estimate decay envelopes in \textbf{c}. \textbf{(f)} For a given allowable phase-slip probability, the enhanced dynamic range of the dual-quadrature readout improves the maximum possible interrogation time. For our particular phase growth profile (inset of \textbf{d}), the improvement is a factor of ${\sim}3.43$.
	} 
	\vspace{-0.5cm}
	\label{Fig3}
\end{figure*}

Arbitrary rotation patterns can be imprinted on the array by shifting all of the atoms by varying distances such that rotations about the $z$-axis with tweezer-resolved phase, $\phi_j$, are applied (Fig.~\ref{Fig2}a). We show the results of time-resolved Ramsey spectroscopy for four different choices of single-site addressing patterns, demonstrating arbitrary, site-revolved, and parallel $\hat{Z}$ rotations (Fig.~\ref{Fig2}b). Such addressing patterns could be used to negate variations in the transition frequency across the array, for instance due to gradients in magnetic field or from the finite differences in tweezer wavelengths as generated by an AOD~\cite{Madjarov2019}. Combining these single-site $\hat{Z}(\phi_j)$ rotations with a series of global $\hat{X}(\pi/2)$ pulses allows for rotations about \textit{any axes}, not just the $z$-axis. As a demonstration (Fig.~\ref{Fig2}c,d), we choose a set of 6 contiguous atoms, initially in the ground state (denoted here as $|{-}Z\rangle$), and rotate them each in parallel into the six states $|{-}Z\rangle, |{+}Z\rangle, |{-}Y\rangle, |{+}Y\rangle, |{-}X\rangle, |{+}X\rangle$, with an average fidelity of 0.984(2) (0.987(2) SPAM-corrected), as determined by state tomography (Methods). The dominant limitations to this value are likely from global drive infidelity and dephasing during the finite shift times.

We note that while here we have demonstrated our protocol on a one-photon optical transition, it could be used to induce a similar effect for two-photon Raman transitions, for instance between hyperfine states~\cite{Levine2022}, assuming the two beams are counter-propagating. Further, the movement-induced phase-shifts employed here rely solely on a relative change in tweezer position, in contrast to alternative techniques that apply additional addressing beams~\cite{Weitenberg2011,Levine2018,Graham2022,Wang2016}, where the phase shift is proportional to a local addressing beam's intensity. While the addressing beam intensity and alignment are prone to drifts on experimental time scales, relative atom movements are ultimately derived from the radiofrequency electronic output of an AWG, which is precise, consistent, and robust. We emphasize our results did not utilize noise-compensating composite pulse sequences and that all data were taken without any system realignments or recalibrations of the atom movements. 

We now demonstrate that access to such robust, high-fidelity, single-site operations can enable enhanced sensing protocols for entanglement-free metrology. In particular, several protocols relying on local control have been proposed for improving the stability of phase-estimation~\cite{Buzek1999,Rosenband2013,Borregaard2013,Li2022} by increasing the dynamic range in which the stochastically evolving laser phase, $\theta$, can be estimated. 

Here we show one such proposal~\cite{Li2022} experimentally, by splitting the array into two sub-ensembles using local addressing to perform Ramsey interferometry simultaneously in two orthogonal bases, $X$ and $Y$, yielding populations $P^{(x)}$ and $P^{(y)}$. While readout in a single basis limits the invertible phase range to $\theta\in[-\pi/2,\pi/2]$, readout in both bases allows this range to be extended unambiguously to $[-\pi,\pi]$ (Fig.~\ref{Fig3}a). Consequently, we can afford a longer Ramsey interrogation time before $\theta$ drifts outside of the invertible range, which would cause a phase-slip error. Note that while the atom number in each quadrature has been halved, this typically does not increase the quantum projection noise (QPN)~\cite{Itano1993} from the dual-quadrature measurement compared to a single-basis measurement (Methods)~\cite{Li2022,Rosenband2013}.

To implement this dual-quadrature readout, we perform Ramsey inteferometry with the addition of a $\hat{Z}(\pi/2)$ rotation to all odd sites in the array before readout (Fig.~\ref{Fig3}b). The resultant oscillations in $P^{(x)}$ and $P^{(y)}$ show a $\pi/2$ phase shift between the even ($X$) and odd ($Y$) sites in the array (Fig.~\ref{Fig3}c). For every repeated measurement (indexed by $j$) at time $t$ we estimate the phase as~\cite{Rosenband2013} 
\begin{align}
\theta_j(t) = \textrm{arg}(z^{(x)}_{j}(t) + i z^{(y)}_{j}(t)),
\label{eq:phaseinversion}
\end{align} 
where \mbox{$z^{(x,y)}_{j}(t) = (2P^{(x,y)}_{j}(t)-1)$} and $\textrm{arg}$ is the argument function. We then calculate the difference, $\delta_j(t)$, of $\theta_j(t)$ from its mean phase (Methods).

We plot the probability distribution $\mathcal{P}(\delta_j(t))$ in Fig.~\ref{Fig3}d, and observe a continuous growth of its standard deviation (STD) $\sigma$ (inset).  We stress that we are interested in the distribution of the laser phase itself, which determines the phase-slip error probability. Hence, we have subtracted off the contribution from QPN to our experimental data shown in the inset of Fig.~\ref{Fig3}d (Methods). We find that the laser phase STD grows with time as a power law $\sigma=\beta t^\alpha$, with $\alpha=0.56(2)$ which we attribute to a power spectral density composed of $1/f$ and white frequency noise. If this standard deviation of the laser phase itself becomes too large compared to the dynamic range, frequent phase-slip errors occur. In Fig.~\ref{Fig3}e we evaluate the phase-slip probability, $\epsilon$, that the phase has exceeded the bounds $[-\pi/2,\pi/2]$ (in emulation of a theoretical single-basis measurement, black dashed lines in Fig.~\ref{Fig3}d), or $[-\pi,\pi]$ (for the dual-quadrature readout); we find that the error probability for the single-basis case quickly becomes substantially larger at shorter interrogation times (Methods).

We further characterize the maximum interrogation time, $T_\textrm{max}(\epsilon)$, for which the phase-slip error probability is still below a threshold $\epsilon$ (Methods). We find that $T_\textrm{max}(\epsilon)$ is significantly increased for the dual-quadrature case (Fig.~\ref{Fig3}f) by a factor of $3.43(13)$, the exact numerical value of which is determined by the phase STD growth rate observed experimentally and is related to the laser noise spectrum (Methods). Such elongation in the attainable interrogation time can be translated directly to enhanced stability in a metrological setting. For example, in a zero dead-time optical clock the stability is proportional to the square-root of $T_\textrm{max}(\epsilon)$, such that we can project an increase in stability by a factor of $\sqrt{3.43}\sim1.8$ for our particular noise profile. This would constitute a practical improvement in phase estimation without increasing the probability of phase-slip errors, a common problem for entanglement enhanced metrology schemes~\cite{Kessler2014,Schulte2020}.

\begin{figure}[t!]
	\centering
	\includegraphics[width=\columnwidth]{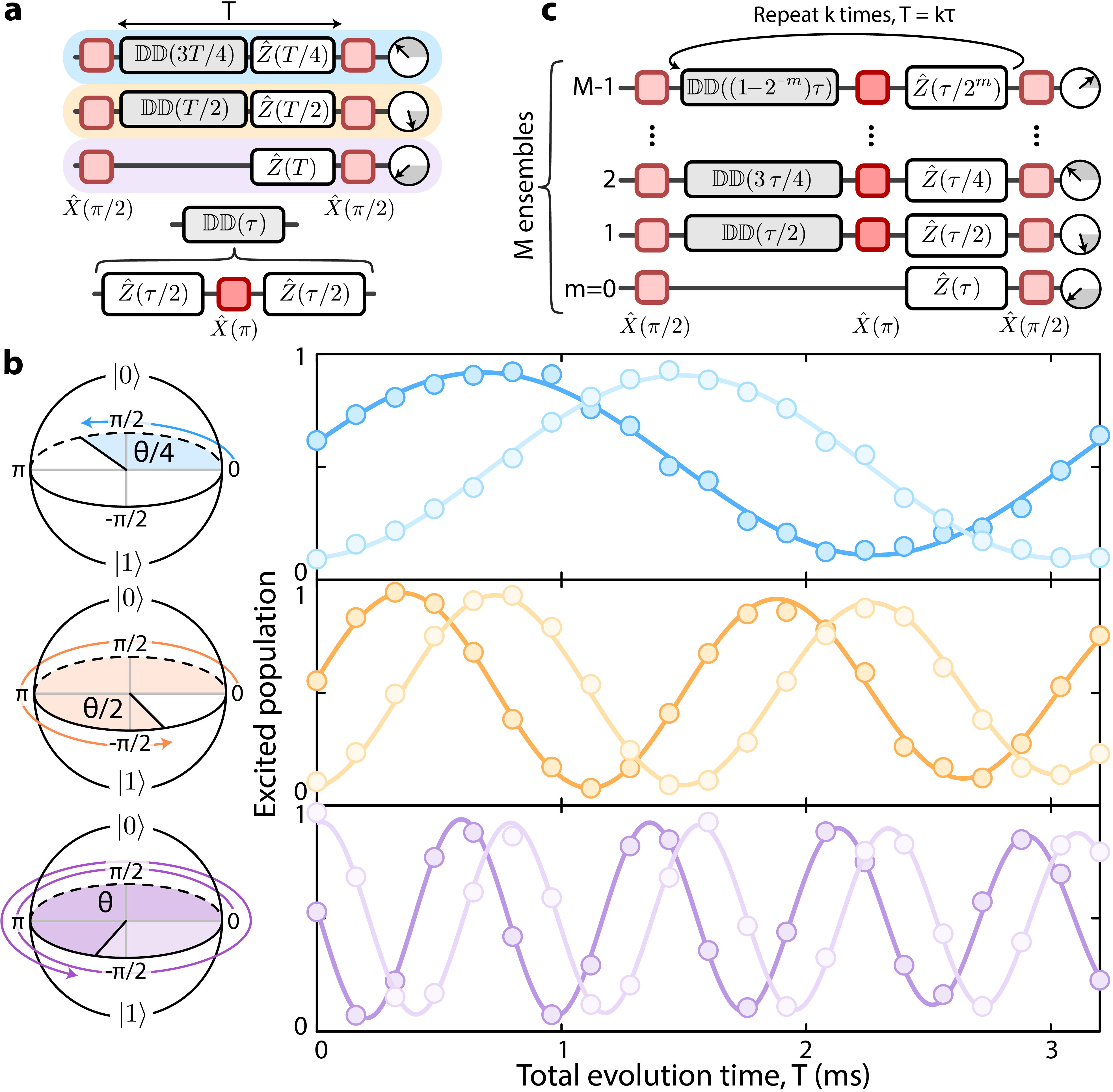}
	\caption{\textbf{Local dynamical decoupling towards optimal metrology.} \textbf{(a)} We split the array into three ensembles, and perform a local dynamical decoupling (DD) sequence such that even though the total Ramsey dark time is $T$, individual ensembles experience different effective evolution times of $T/4, T/2,$ and $T$, respectively. The phase of each ensemble is then measured using dual-quadrature readout. \textbf{(b)} Slower evolving ensembles (those which experience less evolution time) can be used to detect phase-slips in faster evolving ensembles, extending the effective interrogation time of optical clocks. Following the sequence in \textbf{a}, we find the three ensembles evolve at relative rates of 1:1.99(1):4.10(4) with respect to the total evolution time, $T$. The demonstrated scheme in \textbf{a-b} is effective for the case of slow frequency noise where the corresponding noise correlation time is longer than the total evolution time. \textbf{(c)} To handle generic time-dependent noise with shorter correlation times, we envision breaking the total evolution time into $k$ kernels of length $\tau$, each of which is composed of local dynamical decoupling and free evolution. In this way, as long as $\tau$ is shorter than the correlation time of any time-dependent noise affecting the system, the different $M$ ensembles (indexed by $m=0,\cdots, M-1$) can accumulate phase in a correlated manner over the interleaved Ramsey interrogation periods.
 }

	\vspace{-0.5cm}
	\label{Fig4}
\end{figure}

Even greater enhancements in dynamic range, and hence clock stability, could be possible through the use of multiple ensembles with different interrogation times by utilizing fast QND measurements \cite{Bowden2020,Kohlhaas2015} or by
explicitly programming ensembles with different sensitivities to the global laser phase~\cite{Rosenband2013,Borregaard2013}. In the latter of these protocols, the total number of atoms is evenly divided into $M$ ensembles, which are each further subdivided into two sub-ensembles for dual-quadrature measurement. One ensemble is used for normal phase measurement, while for the rest the free evolution time is reduced by factors of $2^{-1},\cdots,2^{1-M}$, or equivalently their effective phase accumulation is reduced by the same amount. If this procedure is performed correctly, the effective ensemble coherence times will then be extended by factors of $2,\cdots,2^{M-1}$, meaning slower evolving ensembles can be used to probe for phase-slips in the fastest ensembles. This then allows for phase estimation over a wider dynamic range beyond $[-\pi,\pi]$, and potentially allows for an improved scaling of the clock stability with atom number~\cite{Rosenband2013} at fixed phase-slip probability (Fig.~\ref{Fig4}a).

As an outlook, we demonstrate a proof-of-principle of local control techniques towards such protocols by performing local dynamical decoupling such that three ensembles experience different effective Ramsey evolution times of $T$, $T/2$, and $T/4$. This is accomplished by inserting local $\hat{X}(\pi)$ pulses (using techniques from Fig.~\ref{Fig2}c) during the evolution at time $T/4$ for the second-fastest ensemble, and time $3T/8$ for the slowest ensemble. Each ensemble is then subdivided further into two sub-ensembles for dual-quadrature readout (Fig.~\ref{Fig4}b). Resultant Ramsey oscillations versus the total evolution time, $T$, show a frequency ratio of 1:1.99(1):4.10(4), very close to the desired 1:2:4 ratio. 

Following this experimental demonstration, we now discuss two limitations (and possible solutions) of this scheme, specifically related to the frequency noise profile and the atom number per ensemble. First, for the simplest case of shot-to-shot noise of laser frequencies that are otherwise constant during the interrogation, our scheme would allow the clock stability to be improved exponentially~\cite{Rosenband2013} by a factor of $\sqrt{2^{M-1}/M}$; the factor of  $\sqrt{1/M}$ stems from increased QPN in the ensemble used for phase estimation and assumes the total number of atoms is distributed uniformly across the $M$ ensembles.
However, for more general time-dependent frequency noise, the situation is more complex, requiring a higher-order pulse sequence~\cite{Rosenband2013}. We propose one such pulse sequence in Fig.~\ref{Fig4}c, by breaking the total evolution time, $T$, into multiple kernels of length $\tau$. Within each kernel, each ensemble experiences a combination of local dynamical decoupling and free evolution, such that the net phase evolution time is $T, T/2,\cdots, T/2^{M-1}$. This scheme could handle noise profiles where the local phase accumulation period, $\tau$, is shorter than the correlation time of the noise. We numerically find that exponential scaling of the maximal interrogation time is then possible up to a saturation point set by the effective decoupling bandwidth (Ext. Data Fig.~\ref{EFig_MultiEns}).

Second, multi-ensemble estimation schemes in general require sufficient atom number per ensemble to be useful~\cite{Rosenband2013}. When the number of atoms per ensemble is limited, quantum projection noise can negate any advantage, i.e. when the error probability in estimating a phase slip by using a slower evolving ensemble exceeds the actual phase slip probability in the fastest evolving ensemble. For the present experimental demonstration with $N\approx6$ per ensemble we do not expect a metrological gain (Ext. Data Fig. \ref{EFig_MultiEns}), but we note that a generalization of our addressing scheme to two dimensional tweezer clock systems~\cite{Young2020} is straightforward. For example, we imagine a realistic scenario of a $10\times20$ atom array with column-by-column control of tweezer positions, such as could be generated with crossed AODs or an AOD combined with a spatial light modulator. In this case, each pair of columns could realize one ensemble with dual-quadrature readout. Finally, we note that such exponential scaling is possible only up to a time-scale where decoherence is dominated by local oscillator noise. Beyond that, interrogation time will be limited by atomic coherence and ultimately by atomic state lifetime~\cite{Bothwell2022}. 

In summary, we have demonstrated arbitrary local rotations for optical transitions through robust phase-sensitive position control in neutral atom arrays, with sub-diffraction limited precision. We have used such rotations to interrogate two atomic ensembles simultaneously for dual-quadrature readout of a Ramsey interferometry signal with demonstrable metrological gain, and have shown a proof-of-principle for controlling many ensembles with variable sensitivity during Ramsey evolution, a key ingredient of proposals for optimal clocks. Further, these methods could be naturally combined with metrologically useful entangled states~\cite{Li2022,Kessler2014,Marciniak2022} to enable simultaneously high sensitivity with a large dynamic range. More generally, our results are an important step towards a fully programmable quantum optical clock based on neutral atoms, which would incorporate quantum computing techniques towards metrological gains, similar to work done with ion trap devices~\cite{Marciniak2022,Schmidt2005} but likely in a more scalable fashion. Such a universal neutral atom clock system would ideally combine arbitrary local rotations, as shown here, with two-qubit entangling operations for optical transitions~\cite{Schine2022}, and mid-circuit readout and reset, which has not been demonstrated so far.

\textit{Note---}During completion of this work we became aware of related work performing local $\hat{Z}$ rotations and studying entanglement-enhanced metrology in an optical tweezer array clock experiment~\cite{Eckner2023}.

\begin{acknowledgements}
We acknowledge useful conversations with Kon Leung, Hannah Manetsch, Su Direkci, and Tuvia Gefen. Further, we thank Jacob Covey for a careful evaluation of our manuscript. We acknowledge support from the Army Research Office MURI program (W911NF2010136), from the Institute for Quantum Information and Matter, an NSF Physics Frontiers Center (NSF Grant PHY-1733907), the NSF CAREER award (1753386), the AFOSR YIP (FA9550-19-1-0044), the DARPA ONISQ program (W911NF2010021), and the NSF QLCI program (2016245). ALS acknowledges support from the Eddleman Quantum Graduate Fellowship. RF acknowledges support from the Troesh postdoctoral fellowship. RBST acknowledges support from the Taiwan-Caltech Fellowship. THY acknowledges support from the IQIM Visiting Fellowship and in part by the NRF (2022M3K4A1094781).
\end{acknowledgements}



 \section*{Data availability}
The data and codes that support the findings of this study are available from the corresponding author upon reasonable request.

\FloatBarrier

\newpage
\FloatBarrier

\bibliography{library_endreslab.bib}
\bibliographystyle{adamref}

\newpage
\clearpage
\section{Methods}
\subsection*{Light-matter interaction Hamiltonian}
After applying the rotating-wave-approximation, the light-matter interaction Hamiltonian considered in this work is given by~\cite{Steck2007}
\begin{align}
\hat{H}=\frac{\hbar |\Omega|}{2}(|0\rangle\langle 1|e^{i k x} + |1\rangle\langle 0|e^{-i k x}),
\label{eq:lightmatter}
\end{align}
where $\Omega$ is the Rabi frequency, $k$ is the global laser wavevector, $x$ is the atomic position along the beam propagation axis, $|0\rangle$ is the ground state and $|1\rangle$ is the excited state. Atom displacements by $\Delta x$ correspond to phase shifts of $\phi=k\Delta x$, as described in the main text. 

Note the choice of the phase for the initial $\hat{X}(\pi/2)$ rotation is a local gauge freedom, and thus all atoms in the array can be said to experience the rotation about the same local axis, e.g. the $x$-axis, despite the spacing between atoms generically not being perfectly commensurate with the driving wavelength. When the atom is shifted, it can be thought of as changing $\hat{X} \rightarrow \hat{X}\cos(\phi)+i \hat{Y}\sin(\phi)$. For the case of only a single global $\hat{X}$ rotation after the movement, this is equivalent to an effective $\hat{Z}(\phi)$ rotation of the quantum state; however, in general if multiple global $\hat{X}$ operations are performed, the equivalence with an effective $\hat{Z}$ rotation breaks down.

\subsection*{Data analysis}
Discrimination between $|0\rangle=|^1S_0\rangle$ and $|1\rangle=|^3P_0\rangle$ is performed by strongly driving the $^1S_0{\leftrightarrow}^1P_1$ transition for $10\ \mu$s, which heats and ejects all atoms in $^1S_0$. (For details on the strontium level structure see Refs.~\cite{Cooper2018,Stellmer2014}). Atoms in $^3P_0$ are then pumped back into $^1S_0$, and imaged with lower power on the $^1S_0{\leftrightarrow}^1P_1$ transition; imaging is performed in 120 ms. 

For all data in the main text, we interleave~\cite{Choi2023} data-taking with feedback to the global $|0\rangle{\leftrightarrow}|1\rangle$ drive frequency every ${\sim}4\ $minutes to counteract slow (${\sim}40\ $minute period) oscillations arising from environmental drifts. $1\sigma$ error bars in the main text are typically smaller than the marker size in all figures; this includes Fig~\ref{Fig1}def, Fig~\ref{Fig3}c, and Fig.~\ref{Fig4}b.

\subsection*{Operation fidelities}
Error modeling suggests that the global $X(\pi)$ fidelity of 0.9956(1) is primarily limited by: measurement errors (see below), finite temperature (the average motional occupation of the atoms along the radial axis is $\bar{n}{\sim}0.2$, leading to a $2\times10^{-3}$ infidelity)), and frequency noise on our laser; the latter of these we believe is also the dominant limitation to our Ramsey coherence time (Fig.~\ref{Fig3}c). The Rabi frequency is ${\sim}2.5$ kHz for all measurements in this work, which allows for fast operations compared to the timescale of decay from $^3P_0$ (${\sim}$550 ms~\cite{Shaw2023}). Measurement errors are dominated by the vacuum-limited atom survival during imaging (0.9995(4)), the imaging fidelity for detecting the presence of an atom (0.9997(2)), and the likelihood of ejecting atoms in $^1S_0$ from the tweezers to perform state discrimination (0.9967(5)) before the final imaging. We note that not all of these measurement errors contribute equally, and their relative importance generically depends on the amount of excited state population in the measured state.

The wavelength of the oscillation, $\lambda_\textrm{osc}$ in Fig.~\ref{Fig1}d is found from fitting the excited state population of shifted atoms with a sinusoid of the form $A\sin(2\pi \Delta x/\lambda_\textrm{osc})+B$, from which we determine the quoted value of $\lambda_\textrm{osc}=699(1)$ nm. Per our independent calibration of distances within the array, we expect there is an additional ${\sim}5$ nm of systematic uncertainty on this measurement. Further systematic uncertainty could arise if the beam propagation is not perfectly coaxial with the array. The crosstalk fidelity of $0.1(2)\%$ is found by fitting the unshifted atom excited state populations in Fig.~\ref{Fig1}d with a sinusoid of the same period as was determined for the shifted atoms; the quoted crosstalk is then the amplitude, $A$, of this sinusoid. We note that we have also checked  for any residual linear phase shifts by repeating this experiment with a global phase shift of $\pi/2$ on the final Ramsey pulse which yields linear sensitivity to small phase shifts. We find the unshifted atoms phase to still be consistent with zero over a range of more than one wavelength.  

To determine the fidelity of arbitrary local rotations (Fig.~\ref{Fig2}c), we perform quantum state tomography by reading out the produced states in the $x$-, $y$- and $z$- bases by rotating the state with a global $\pi/2$ pulse of a given global laser phase as necessary. The fidelity is then estimated with $F = \langle\psi_{\textrm{target}}|\rho|\psi_{\textrm{target}}\rangle$, where $\rho$ is the experimental state determined by quantum tomography. Due to the choice of the six arbitrary rotations, the fidelity estimation coincides with the population of the excited state or the ground state along the corresponding axis. For instance, the fidelity of $|{+}Y\rangle$ is determined by the population of the excited state when the prepared state is rotated and measured in the $y$-basis. In order to access the intrinsic infidelity induced by arbitrary local rotations, we extract SPAM errors and correct them. SPAM sources are dominated by the same detection infidelities as for the global $\hat{X}(\pi/2)$ fidelity (see above), and the finite SPAM-corrected readout $\pi/2$ pulse fidelity of 0.9982(4). 

\subsection*{Estimating laser phase prior width}
Here, we detail how we find and isolate the prior width of the laser phase distribution from the experimentally measured values. We note that while the phase slip probability depends only on this prior width, the measured distribution is further affected by quantum projection noise (QPN). 
The QPN itself is a function of the fraction of excited atoms measured in the Ramsey sequence.  
One thus finds that the relative contribution of the QPN term varies with the central phase of the laser and the prior width of the phase distribution. Assuming a system of $N=20$ atoms, we plot the calculated phase distribution width as a function of central laser phase $\bar{\theta}$, but with the underlying prior phase distribution having a width of zero (Ext. Data Fig.~\ref{EFig_QPN}(a)). We repeat this calculation for both single and dual quadrature phase estimation.
We note that these estimators are affected slightly differently by QPN. Specifically, while the optimal single quadrature working point in terms of minimal QPN is around a mean phase of 0 (which corresponds to measurement with an excitation fraction of 0.5), we find that the optimal working point for a dual-quadrature estimation is around a mean phase of $\pi/4$ (corresponding to the two quadratures having excitation fractions of 0.85), though we note the dual-quadrature value is relatively flat over the entire bandwidth. For longer interrogation times, the prior width grows as a power law which depends on the laser noise spectrum. The contribution from projection noise is thus in general time-dependent.

To isolate this effect and learn the true laser phase distribution as a function of interrogation time, we first calculate the total observable width including QPN, $\sigma_{tot}$, for a range of prior widths $\sigma_{\delta}$ at a given laser central phase $\bar{\theta}$. This is done by sampling random phases from a normal distribution, followed by sampling the observed phase from a binomial random process representing the projection uncertainty. Repeating this process over a million draws we obtain the observed distribution as a function of the prior width (see Extended Data Fig. \ref{EFig_QPN}(b)). We then invert the function to obtain $\sigma_{\delta}$ ($\sigma_{tot}$) and interpolate the latter to find the prior width at the given measured $\sigma_{tot}$.


\subsection*{Fitting the phase-slip probability}
To find the phase deviation from the mean phase for a given shot, we fit the Ramsey oscillations in Fig.~\ref{Fig3}c with a decaying sinusoid, and at each time define the fitted mean populations as $\bar{P}^{(x)}(t)$ and $\bar{P}^{(y)}(t)$. These populations are inverted via Eq.~\ref{eq:phaseinversion} into the mean phase, $\bar{\theta}(t)$, and finally we calculate the phase deviation from the mean as \mbox{$\delta_j(t)=\textrm{mod}(\theta_j(t)-\bar{\theta}(t),\pi)$}. 

For the phase-slip probability (Fig.~\ref{Fig3}e), we fit the probability densities $\mathcal{P}(\delta_j(t))$ with $G(\mathcal{P})$, where $G$ is a Gaussian distribution folded into the range $[-\pi,\pi]$. This fit provides an estimation of the true standard deviation, $\sigma$,  of the $\delta_j(t)$ distribution. With this in hand, for a given half-dynamic range, $B$, we then find the phase-slip probability, $\epsilon$, as 
\begin{align}
\epsilon=2\int_B^\infty G(\mathcal{P})d\mathcal{P}=\textrm{erfc}\Big(\frac{B}{\sqrt{2}\sigma}\Big),
\label{eq:phaseslipprobability}
\end{align}
where $\textrm{erfc}$ is the complementary error function. Note that here $B=\pi/2$ corresponds to a single-basis measurement while $B=\pi$ corresponds to a dual-quadrature measurement.

In order to calculate the maximal interrogation time, we first find the laser phase prior width (as described in the previous section) and then fit the time-resolved profile of $\sigma(t)$, as $\sigma(t)=\beta t^\alpha$. We find $\beta=\pi\times0.119(6)$ and $\alpha=0.56(2)$. The growth of $\sigma$ over time can then be used to predict the Ramsey decay envelope, $C$, for different choices of $B$, as $C=e^{-\sigma^2/2}$. In Fig.~\ref{Fig3}c, we show this envelope estimation for the cases of $B=\pi/2$ and $B=\pi$ (orange and green dashed lines, respectively).

We then analytically calculate the maximum interrogation time at a fixed phase-slip error probability, $\epsilon$, from Eq.~\ref{eq:phaseslipprobability} as
\begin{align}
T_{\textrm{max}}(\epsilon)=(B/(\sqrt{2}\beta \textrm{erfc}^{-1}(\epsilon)))^{1/\alpha}.
\end{align} 

\subsection*{Limits in multi-ensemble metrology}

To study the possible limitations of the multi-ensemble scheme we simulate stochastic phase evolution of a local oscillator with $1/f$ frequency noise, whose overall power sets a characteristic single ensemble $1/e$ Ramsey coherence time $\mathrm{T}_{\mathrm{LO}}$. We numerically find the maximal interrogation time at a fixed phase slip probability $\epsilon$ (here we use $\epsilon=5\cdot10^{-3}$) with increasing ensemble number $M$ by iteratively correcting for phase slips as described in Ref. \cite{Rosenband2013}. We repeat this calculation for different dynamical decoupling block lengths $\tau$.
In Ext. Data Fig.~\ref{EFig_MultiEns}a we plot the results for up to $M=9$ ensembles, assuming infinite atom number. We find that adding more ensembles indeed enables exponential scaling of the interrogation time up to a saturation point set by the effective dynamical decoupling bandwidth (expressed in terms of $\tau/\mathrm{T_{LO}}$).

We further study the effect of quantum projection noise on the efficacy of the scheme in the case of low atom number per ensemble. For the optimal dynamical decoupling sequence found previously, we vary the number of atoms per ensemble $N$ and repeat the calculation, which is now affected by quantum projection noise. Specifically, the use of slower evolving ensembles with limited atom number for the iterative correction of phase slips in the fastest evolving ensembles is prone to errors due to the increased variance in such estimation.  
For a small number of atoms per ensemble, this negates any advantage. However, we note that $N\simeq20$ atoms per ensemble suffice for efficient operation of the scheme, with the interrogation times and noise strength tested here.

\FloatBarrier
\clearpage
\newpage
\setcounter{figure}{0}

\captionsetup[figure]{labelfont={bf},name={Ext. Data Fig.},labelsep=bar,justification=raggedright,font=small}

\section*{Extended Data Figures}

\begin{figure}[ht!]
	\centering
	\includegraphics[width=\columnwidth]{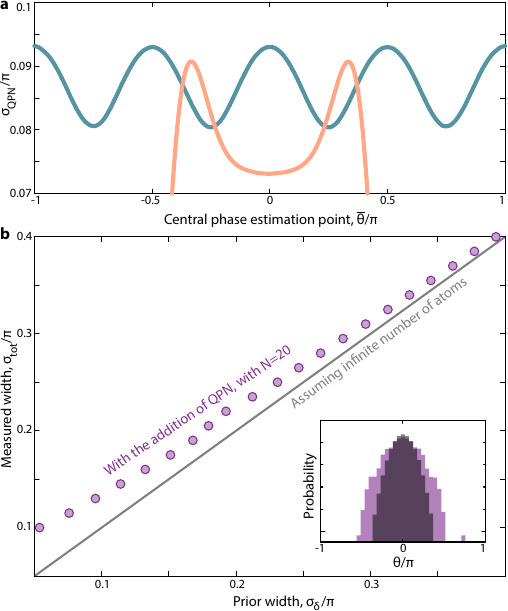}
\caption{\textbf{Quantum projection noise in a dual-quadrature measurement.} 
\textbf{a,} Added standard deviation due to quantum projection noise (QPN) for phase estimation around different average phases $\bar\theta$, plotted for $N=20$ atoms utilized in a single-quadrature (orange) or dual-quadrature (green, 10 atoms per quadrature) measurement. The added QPN varies with the phase the measurement is taken at; thus as the prior width of the phase distribution grows over time, and a broader range of phases is sampled, the QPN will vary. \textbf{b,} To learn the prior width from the measured width we sample random phases from a normal distribution, followed by sampling the observed phase from a binomial random process representing the projection uncertainty (inset). We use the sampled distributions for the dual-quadrature estimator to calculate the width including QPN for a range of prior laser widths. We then invert this function and interpolate if needed, to find the prior width for any measured width.}  

	\label{EFig_QPN}

 \end{figure}

\begin{figure}[ht!]
	\centering
     \includegraphics[width=\columnwidth]{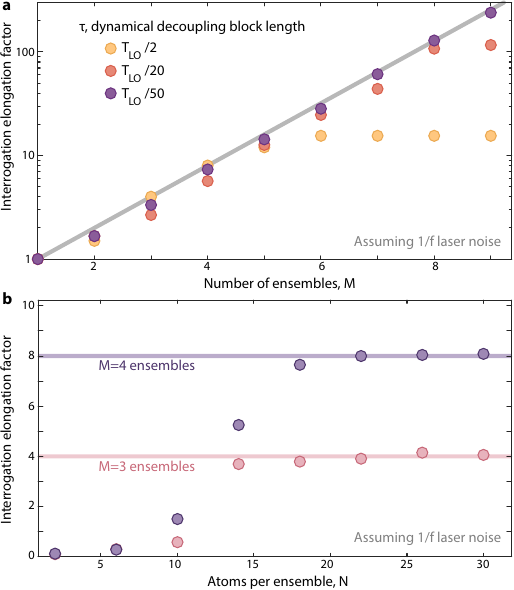}
\caption{\textbf{Limits in multi-ensemble metrology.}  \textbf{a,} Asymptotic scaling of the extended interrogation time factor as a function of the number of ensembles M employed. We numerically calculate the maximal interrogation time at a fixed phase slip probability for different dynamical decoupling block lengths $\tau$, in terms of the local-oscillator coherence time $\mathrm{T}_\mathrm{LO}$, assuming $1/f$ frequency noise and infinite atom number. The addition of more ensembles enables exponential scaling of the maximal interrogation time (solid line marks $2^{M-1}$) up to a saturation point set by the effective decoupling bandwidth. The latter can be extended by reducing the block length while maintaining a sufficiently high Rabi frequency with respect to the fast noise frequency. \textbf{b,} For the optimal decoupling sequence found in \textbf{a}, we plot the extended interrogation time as a function of the number of atoms per ensemble $\mathrm{N}$. For a small number of atoms per ensemble, quantum projection noise results in an enhanced rate of false positive indication of a phase slip, negating any advantage. We find that $\mathrm{N}\simeq20$ atoms per ensemble suffice for efficient operation.}
	\label{EFig_MultiEns}

 \end{figure}

\FloatBarrier
\newpage

\end{document}